\documentclass[12pt,preprint]{aastex}

\begin{document}

\title{CIRCUMSTELLAR Na\,I AND Ca\,II LINES IN TYPE IIP SUPERNOVAE 
AND SN~1998S}

\author{N. N. Chugai}
\affil{Institute of Astronomy, RAS, Pyatnitskaya 48, 119017 Moscow, Russia}
\email{nchugai@inasan.ru}

\and

\author{V. P. Utrobin}
\affil{Institute of Theoretical and Experimental Physics, Moscow 117218,
B.~Cheremushkinskaya St. 25}
\email{utrobin@itep.ru}

\begin{abstract}

We study a possibility of detection of circumstellar 
absorption lines of Na\,I D$_{1,2}$ and Ca\,II H,K in spectra of 
type IIP supernovae at the photospheric epoch. The modelling shows that 
the circumstellar lines of Na\,I doublet will not be seen in type 
IIP supernovae for moderate wind density, e.g., characteristic of 
SN~1999em, whereas rather pronounced Ca\,II lines with P Cygni 
profile should be detectable. A similar model is used to describe 
Na\,I and Ca\,II circumstellar lines seen in SN~1998S, type IIL with 
a dense wind. We show that line intensities in this supernova 
are reproduced, if one assumes an ultraviolet excess, which is caused 
primarily by the comptonization of supernova radiation in the 
shock wave.

\end{abstract}

\section{Introduction}

Type IIP supernovae (SN IIP) presumably originate from stars with 
initial masses in the range of $9-25~M_{\odot}$ (Heger et al. 2003).
Prior to the explosion a pre-SN IIP is usually a red supergiant 
(RSG) (Grasberg et al. 1971) that presumably loses matter in a form of 
a slow dense wind. It would be reasonable to assume that the mass loss rate 
should correspond to RSG with the initial mass characteristic of 
SN IIP, i.e., $\sim(1-10)\times10^{-6}~M_{\odot}$ yr$^{-1}$ 
(Chevalier et al. 2006). However, it is not yet clear that this is 
always the case. There is an opinion that massive RSG 
($10-20~M_{\odot}$) during the last $10^4$ yr before the gravitational
collapse of iron core could
lose matter in the form of superwind with the rate of 
$\sim10^{-4}~M_{\odot}$ yr$^{-1}$ owing to pulsation instability 
(Heger et al. 1997). On the other hand, for type IIP SN 1999em with
the known mass of pre-SN of $\approx 20~M_{\odot}$ the mass loss 
rate is $\dot{M}\sim10^{-6}~M_{\odot}$ yr$^{-1}$ (Chugai et al. 2007),
which is lower than not only the pulsation mass loss rate but also
the value of $\sim8\times10^{-6}~M_{\odot}$ yr$^{-1}$, predicted 
by the phenomenological relation of Nieuwenhuijsen and de Jager (1990) 
for a RSG with the same main sequence mass. This 
disparity emphasises the significant uncertainty in the problem of 
the mass loss by pre-SN IIP. To compose a more clear picture one needs 
to obtain sufficiently large sample of SN IIP with the estimated 
density of the circumstellar (CS) gas.

At present the mass loss rate by pre-SN IIP is estimated from 
radio and X-ray emission originated from a shock interaction 
between supernova ejecta and the wind (Chevalier 1982; Pooley et al. 
2002), perhaps, with the more reliable estimates based on X-ray data. 
For SN 1999em, SN 1999gi, SN 2004dj, and SN 2004et mass 
loss rates recovered from X-ray data are confined in the range 
of $(1-2.5)\times10^{-6}~M_{\odot}$ yr$^{-1}$ (Chevalier et al. 
2006; Rho et al. 2007), whereas for SN 2006bp the value of
$\sim10^{-5}~M_{\odot}$ yr$^{-1}$ is obtained (Immler et al. 2007).
Recently another method based on the high velocity components of 
H$\alpha$ and He\,I 10830 \AA\ lines is proposed which in case of SN 1999em 
results in the estimate of $\approx10^{-6}~M_{\odot}$ yr$^{-1}$
(Chugai et al. 2007).

Here we investigate a more direct diagnostic tool for estimating
the wind density based on 
the observation of CS absorption lines of Na\,I D$_{1,2}$ and 
Ca\,II H,K against the luminous supernova photosphere. Up to now 
these lines have been confidently detected only in type IIL 
SN 1998S (Bowen et al. 2000). A search for these lines in SN IIP 
has not yet been performed, although at present the search for 
CS lines in SN Ia is actively carrying out
 (Patat et al. 2007a; Patat et al. 2007b). 
In the case of SN 1998S the wind density according to high X-ray and 
radio luminosity is large and corresponds to the mass loss rate of
$\sim2\times10^{-4}~M_{\odot}$ yr$^{-1}$ (Pooley et al. 2002). 
By this reason it is not yet clear whether Na\,I and Ca\,II lines 
could be observed in SN IIP in which case the mass loss rate is 
significantly lower than in SN 1998S.

In the present paper we study the formation of Na\,I and Ca\,II lines in 
the RSG wind after the SN IIP explosion and the use of these lines 
for the diagnostics of the wind density.
We start with the description of the model (section 2), compute 
the ionization of Na\,I and Ca\,II in the wind before and 
after the explosion, and then present model profiles of CS lines 
of Na\,I 5890 \AA\ and Ca\,II 3934 \AA\ for typical wind densities 
(section 3). We then apply our model to the explanation of 
circumstellar lines of Na\,I 5890 \AA\ and Ca\,II 3934 \AA\ in 
SN 1998S and discuss conditions for which these lines have the 
observed intensities (section 4). In conclusion we consider 
the possibility of detection of CS lines and discuss factors that might 
lead to the deviations of line intensities from model results.

\section{Model}

We consider below a spherically-symmetric stationary wind with 
the density $\rho=w/(4\pi r^2)$ and velocity $u$, in which SN IIP 
explodes. It is convenient to deal with the dimensionless parameter 
$\omega$ defined by the relation $w=6.3\times10^{13}\omega$ g cm$^{-1}$; 
the values $\omega=1$ corresponds to the mass loss rate of
$10^{-6}(u/10\,\mbox{km s}^{-1})~M_{\odot}$ yr$^{-1}$.
Before the supernova explosion the wind hydrogen is neutral, whereas 
Na and Ca could be singly ionized by the RSG radiation. The major 
ionizing factor is the chromospheric radiation of RSG. 

A general idea about the intensity of the chromospheric radiation 
of pre-SN provides the galactic RSG $\alpha$ Ori (Betelgeuse). According to 
the data obtained with {\em IUE} (Rinehart et al. 2000), the fluxes in 
1250-1750 \AA\ and 1900-3200 \AA\ bands are 
$(4-6)\times10^{-11}$ and $(2-3)\times10^{-9}$ erg cm$^{-2}$ s$^{-1}$,
respectively. For the power law approximation $f_{\lambda}\sim \lambda^q$
these fluxes are reproduced with $q=5$. The absolute monochromatic 
luminosity is determined adopting the standard distance of 131 pc
for $\alpha$ Ori. To calculate ionization of metals in the 
pre-SN wind, we solve numerically a time-dependent ionization balance 
taking into account the wind expansion with the velocity 
$u=15$ km s$^{-1}$ 
assuming the same ultraviolet luminosity as for Betelgeuse. Metals
Mg, Si, and Fe, dominating in the electron number density, are treated 
as a single element with the relative abundance of $10^{-4}$
with respect to the hydrogen and with the ionization potential of 7.9 eV. 
Time dependent ionization is solved on the time interval of $10^5$ yr.
The wind temperature is assumed to be equal to the local 
radiation temperature $T=T_{\rm s}W^{0.25}$, where $W$ is the dilution 
factor, while $T_{\rm s}=3900$~K is the effective temperature of the 
RSG, which corresponds to the luminosity of $10^5~L_{\odot}$ and 
the radius of $700~R_{\odot}$.

The calculated ionization fractions of Na\,I and Ca\,II in the pre-SN 
wind are used then as initial conditions for the calculations of the 
time-dependent ionization of these ions after the supernova explosion 
(cf. Chugai 2008). The high initial supernova luminosity with the 
temperature $\geq10^5$~K results in the strong ionization of 
hydrogen in the wind, which then has not enough time to recombine during 
the considered period of 50 days. We therefore adopt the complete ionization
of wind hydrogen. The metal ionization is calculated with the fixed 
 wind electron temperature of $3\times10^4$~K, the log-average between 
 extreme values of $10^4$ K and $10^5$ K (Lundqvist and Fransson 1988).
The supernova bolometric luminosity and the velocity at the photosphere 
are adopted to be equal to those of SN~1999em (Utrobin 2007). To describe 
ultraviolet spectrum, we introduce a reduction factor for the black body 
radiation; this factor depends on the wavelength and time according to 
the evolution of the ultraviolet spectrum of SN 1987A (Pun et al. 1995). 

To compute line profiles, we consider the wind outside the shock wave,
which coincides with the contact surface at the boundary between the supernova ejecta and
the wind in the thin shell model (Chevalier 1982). The evolution of the radius 
of this shell is calculated numerically assuming the ejecta mass of 
$18~M_{\odot}$ and the kinetic energy of $1.3\times10^{51}$ erg, close 
to the parameters of SN~1999em (Utrobin 2007). The density distribution of  
the supernova envelope is set as a combination of internal plateau for $v<v_0$,  
external power law drop $\rho\propto v^{-9}$, and outer cutoff at 
$v=v_{\rm b}$. This cutoff is related with the 
shock wave breakout and the transition from adiabatic to radiative regime 
(Grasberg et al. 1971). The adopted boundary velocity is 
$v_{\rm b}=15000$ km s$^{-1}$ in accordance with radial velocities 
in the blue wing of H$\alpha$ absorption in early spectra of 
normal SN IIP, e.g., SN~1999em (Leonard et al. 2002a) and 
SN~1999gi (Leonard et al. 2002b). Note, the adopted boundary velocity is 
qualitatively consistent with the hydrodynamic modelling of SN~1999em 
which gives the value  $v_{\rm b}=13400$ km s$^{-1}$ (Utrobin 2007).
 
\section{Results}

According to our model calculations the metals in the pre-SN wind 
turn out strongly ionized within considered zone $r<10^{18}$ cm. 
Fractions ($y$) of Na\,I/Na and Ca\,II/Ca as a function of radius 
are shown in Fig. 1a for the wind density parameter 
$\omega=1$ and 10. In the internal wind zone $r<10^{16}$ cm 
one gets $y(\mbox{Na\,I})\sim 10^{-3}-5\times10^{-2}$ and 
$y(\mbox{Ca\,II})\sim 0.1-1$; at the larger distance the value 
$y(\mbox{Ca\,II})$ is lower by an order of magnitude.
The supernova explosion results in the significant enhancement of 
the metal ionization. The distribution of the relative concentrations 
of Na\,I/Na and Ca\,II/Ca in the wind on day 50 after the explosion is 
presented in Fig. 1b for the same density parameter values $\omega=1$ and 10.
The Na\,I ionization is strong everywhere, while Ca\,II is strongly 
ionized only in the outer zone where recombination is suppressed 
because of the low density.

At first glance the setting of the wind conditions for a single moment 
is nonsense, because this does not take into account light travel 
effect. In fact, however, the photon absorption is determined by the age of 
supernova $t_1$, when the photons were emitted by the photosphere. 
Indeed, at the moment $t_1+r/c$, when photon packet attains the point $r$, 
where they can be absorbed, the state of the wind is determined by the 
radiation emitted in the interval $0<t<t_1$ independent of the $r$ value. 
Moreover, for the observer at the distance $D$ the 
moment of detection of this photon packet, $t_0=t_1+r/c+(D-r)/c-D/c=t_1$, 
coincides with the supernova age $t_1$. To summarize, when only absorption 
is considered the light travel effects do not present explicitly. 
This statement is true with the accuracy of $u/c\ll1$, where 
$u$ is the wind velocity. For photons scattered at the radius $r$ 
by angle $\theta$ towards the observer the detection moment 
$t_0=t_1+r(1-\cos\,\theta)/c>t_1$ is larger than the supernova age, 
i.e., the light travel effects should be taken into account in this case 
(see below).

The wind optical depth $\tau$ in Na\,I 5890 \AA\ and Ca\,II 3934 \AA\ lines
outside the shock wave on days 15 and 50 is 
given in Fig. 2 for the same wind density and temperature as above 
and the turbulent velocity of 2 km s$^{-1}$. Interestingly, 
the optical depth of Ca\,II 3934 \AA\ is contributed primarily by 
the inner region $r<6\times10^{15}$ cm, while the Na\,I 5890 \AA\ 
line by the region around $r\sim10^{16}$ cm. In both lines 
$\tau$ grows with time. Note, the optical depth of the Na\,I 5890 \AA\ line 
for the wind density $\omega=1$, characteristic of SN~1999em, is small 
even on day 50 ($\tau\sim0.05$), whereas the optical depth 
 in the Ca\,II 3934 \AA\ line is 
large not only on day 50 but on day 15 as well. Only for very dense 
wind $\omega\approx10$ the optical depth in the Na\,I 5890 \AA\ line 
is large ($\tau>1$) at the late photospheric phase ($t\sim50$ d).

The obtained distributions of number density of Ca\,II and Na\,I in the 
wind permit us to compute line profiles of CS lines via direct integration 
of the equation of radiation transfer. The source function is 
determined in the escape probability approximation 
assuming complete frequency redistribution
\begin{equation}
S=\frac{\beta WI_{\rm c}}{\beta+(1-\beta)\epsilon}\,,
\end{equation}
where $W$ is the dilution factor, $I_{\rm c}$ is the photosphere brightness,
$\beta$ is the Sobolev escape probability, $\epsilon$ is the photon 
destruction probability. In the resonance Na\,I line the scattering is 
conservative ($\epsilon=0$), while in the 
Ca\,II 3934 \AA\ line we take into account photon destruction due to 
the fluorescence in the infrared triplet lines ($\epsilon=0.068$). 
Light travel effects in the profile computations are taken into account  
approximately by discarding the region for which the 
light delay is greater than the supernova age. The 
occultation by the photosphere and the resonance 
scattering by Na\,I and Ca\,II in the supernova atmosphere are taken into 
account. To this end 
we assume that the inner scattering zone of the supernova envelope 
is bounded by the velocity of 0.8 of the maximal velocity. 
The wind velocity is set to be 15 km s$^{-1}$, the value 
found for Betelgeuse (Huggins et al. 1994). The turbulent velocity is 
set to be 2 km s$^{-1}$. This value is based on the turbulent velocity in 
the wind of Betelgeuse $v_{\rm t}\approx 1$ km s$^{-1}$ and on the 
estimate of the velocity dispersion due to the radiative acceleration 
after the supernova explosion 
\begin{equation}
u=\frac{k_{\rm T}E_{\rm r}}{4\pi r^2c}=0.9E_{\rm r,49}r_{16}^{-2}\;\;
\mbox{km s$^{-1}$}\,,
\end{equation}
where $k_{\rm T}=0.34$ cm$^2$ g$^{-1}$ is the Thomson opacity, 
$E_{\rm r}$ is the radiated energy, $r$ is the radius; numerical 
indices indicate units in $10^{49}$ erg and $10^{16}$ cm,
respectively. This relation shows that in the region 
$r\sim (0.4-1)\times10^{16}$ cm, which contributes mostly to the 
optical depth of Ca\,II line (Fig. 2), one obtains for 
$E_{\rm r,49}\approx0.5$ at about day 40 the velocity dispersion 
of $\approx1-2$ km s$^{-1}$, so that the total dispersion in the wind 
is about 2 km s$^{-1}$. The Doppler width is calculated in a standard 
way using turbulent and thermal velocity.

The calculated line profiles of Na\,I 5890 \AA\ and Ca\,II 3934 \AA\ 
on days 15 and 50 for $\omega=1$ and 10 are plotted in Fig. 3.
The profiles are convolved with the Gaussian instrumental profile 
FWHM=10 km s$^{-1}$ to mimic the typical spectral resolution.
Calculated line profiles have strong emission component, which is 
consistent with its formation in the inner wind zone in which 
light travel effects are not pronounced. Note, the emission component 
may serve as a signature that the line forms in the wind, not in the 
interstellar medium. It should be emphasized that Ca\,II line is 
strong on days 15 and 50 even for moderate density ($\omega=1$) 
whereas the Na\,I 5890 \AA\ line gets noticeable only for 
rather dense wind $\omega\approx10$ and on the late stage 
$t\sim 50$ d. Equivalent width of the Ca\,II 3934 \AA\ absorption 
grows with $\omega$ approximately as 
\begin{equation}
W_{\lambda}\approx0.13(1+0.385\lg\,\omega)\, \mbox{\AA}. 
\end{equation}
This relation can be used for a rough estimate of the wind density 
in SN IIP using the CS absorption Ca\,II 3934 \AA\ around day 50.

\section{Type IIL supernova 1998S}

It is tempting to apply our model to the interpretation of 
CS lines of Na\,I and Ca\,II, detected in spectra of SN 1998S. 
This supernova belongs to bright variety of SN IIL; in fact this is 
a close analogue of SN 1979C (Liu et al. 2000). According to X-ray 
data the wind around SN 1998S is characterized by the mass loss rate 
of $(1-2)\times10^{-4}~M_{\odot}$ yr$^{-1}$ assuming wind velocity of
10 km s$^{-1}$ (Pooley et al. 2002). The corresponding wind density 
parameter is $\omega\sim200$. The extrapolation of results obtained 
above suggests that strong circumstellar Ca\,II and Na\,I lines 
should be present 
in the spectrum of this supernova with the equivalent width of 
Ca\,II 3934 \AA\ of $>0.2$ \AA. 

Indeed high resolution spectra of SN~1998S show CS lines of 
Na\,I D$_{1,2}$ doublet with the growing intensity between 
days 20 and 39 after the outburst (Bowen et al. 2000). In the 
3934 \AA\ band on day 39 the spectrum shows similar 
CS component of Ca\,II 3934 \AA. Despite the expectation the circumstellar
Ca\,II 3934 \AA\ line has a moderate intensity with the equivalent width 
of 0.1 \AA\ and a relative depth of 0.5.
 
To reproduce CS lines in SN~1998S, we use the model applied above for SN IIP 
with the following modifications. The bolometric light curve and the effective
temperature evolution correspond to SN~1998S (Fassia et al. 2000), 
while the wind density is $\omega=200$. The 
adopted wind velocity is 40 km s$^{-1}$ (Fassia et al. 2001); 
the turbulent velocity is assumed to be 5 km s$^{-1}$, higher than for 
SN IIP, because the radiated energy of SN~1998S is 2-3 times larger than 
that for SN IIP on day 40. The envelope mass of SN 1998S can be estimated 
from the following considerations. The mass of mixed metal core in 
the velocity range $v\leq 3650$ km s$^{-1}$ is about $4~M_{\odot}$ 
(Fassia et al. 2001). The major envelope mass is confined within 
the velocity of 5000 km s$^{-1}$ (Fransson et al. 2005). Assuming 
homogeneous density distribution we find that the total mass is 
$M=10~M_{\odot}$. Since the density should fall towards higher velocities, 
the mass should be $M<10~M_{\odot}$; we adopt $M=8~M_{\odot}$. 
The kinetic energy is taken the same as in SN~IIP, i.e., 
$E=1.3\times10^{51}$ erg. Note, the uncertainty in mass and energy only weakly 
affects the final results. 

Preliminary modelling shows that for the black body continuum 
CS absorption lines turn out too strong. The natural mechanism for 
the suppressing of line intensity could be an ultraviolet excess in the 
SN 1998S spectrum. There are two reasons for the emergence of 
this excess: Compton scattering on hot electrons of the forward shock 
wave (Fransson 1984) and intrinsic emission of the gas in the 
shock wave. We consider, therefore, two options for the supernova 
spectrum: (1) black body spectrum and (2) black body continuum with 
the ultraviolet excess $F_{\nu}\propto \nu^{-3}$ in the region 
$\lambda<2000$ \AA. Integrated flux of the ultraviolet excess 
makes up the fraction $\eta$ relative to the black body
flux $\sigma T^4$. The first option corresponds to $\eta=0$, while 
the second to $\eta>0$. We find that the optimal value of the 
ultraviolet excess is $\eta=0.06$.

The results of computations of the optical depth in 
Na\,I 5890 \AA\ and Ca\,II 3934 \AA\ lines on days 20 and 39 
for $\eta=0$ and  $\eta=0.06$ are presented in Fig. 4. 
The line intensity increases with time and Ca\,II line is stronger than 
Na\,I line in the same way as for SN~IIP. In the case $\eta=0$ CS lines 
are stronger than for $\eta=0.06$, which is a natural outcome of 
a more stronger ionization in the latter case. Moreover, 
for $\eta=0$ the intensities of 
Ca\,II and Na\,I lines differ stronger than for $\eta=0.06$ since 
in the latter case the ultraviolet excess ionizes Ca\,II relatively 
stronger than Na\,I, which results in the equalizing of 
 Na\,I and Ca\,II concentrations. Observed CS 
Na\,I 5890 \AA\ and Ca\,II 3934 \AA\ lines in SN~1998S have moderate 
intensities and differ weakly. By these signatures the case 
$\eta=0.06$ should be preferred compared to $\eta=0$.

The above said is illustrated by Fig. 5 that shows calculated profiles 
of Na\,I 5890 \AA\ and Ca\,II 3934 \AA\ in SN~1998S on days 20 and 39 
in the case of $\eta=0$ (Fig. 5a,b) and $\eta=0.06$ (Fig. 5c,d).
The case with ultraviolet excess describes observed CS lines of
Na\,I 5890 \AA\ and Ca\,II 3934 \AA\ in SN~1998S (cf. Bowen et al. 
2000, their Fig. 4) much better than the case $\eta=0$ that predicts 
unacceptably strong lines on day 39. We conclude that the moderate 
intensity of 
Na\,I 5890 \AA\ and Ca\,II 3934 \AA\ and their resemblance 
are related with the presence of the ultraviolet excess 
compared to the black body radiation in the spectrum of SN~1998S.

To what extent the required ultraviolet excess is consistent with 
the shape of ultraviolet spectrum on day 30 taken from 
{\em HST} data (Fransson et al. 2005) and with comptonized 
black body spectrum? We computed the comptonized spectrum in the 
single scattering approximation (Rephaeli \& Yankovitch 1997) 
adopting parameters of SN 1998S on day 30, i.e., the radiation 
temperature of 9440 K, electron temperature in the forward shock of 57 keV, 
for the Thomson optical depth of the forward shock $\tau_{\rm T}=0.13$. 
The computed spectrum in comparison with the required ultraviolet excess 
is shown in Fig. 6. We show there also the observed spectrum corrected for 
the reddening $E(B-V)=0.26$, which is slightly higher than the value 
0.22 adopted by Fassia et al. (2000), but still within reported
 uncertainties. The figure shows reasonable agreement between 
 the required ultraviolet excess and both observed and computed 
 comptonized spectrum. Yet it should be noted that on day 39 the 
 computed ultraviolet comptonized flux is weaker by 1.3 times 
 than the required ultraviolet excess. We suggest that this deficit 
 is covered by the thermal radiation of the shock.
 
It was mentioned already that the characteristic signature of model 
CS lines is the presence of an emission component. We note that 
the comparison of the observed Na\,I D$_{1,2}$ profiles on days 
20 and 39 (Bowen et al. 2000) indeed shows the presence of emission 
component on day 39. This is an additional argument in favour of the CS origin 
of the blue component of Na\,I D$_{1,2}$ blend in SN 1998S.

\section{Conclusion} 

The primary goal of the paper was to construct the model of the formation 
of Na\,I and Ca\,II CS lines in the wind around SN IIP in a hope 
to use them for the wind diagnostics.
The modelling shows that lines of Na\,I doublet will not be seen in SN~IIP 
spectra for moderate wind density, $\omega\sim 1$, but will be detectable 
at late photospheric stage $t\geq50$ d in the case of the dense wind 
$\omega\sim 10$. Yet the Ca\,II lines will be seen even in the case of 
a rarefied wind $\omega<1$ and therefore they are especially advantageous 
for the detection of the wind around SN IIP. We predict that the 
spectrum with the resolution of $\approx10$ km s$^{-1}$
of a normal SN IIP at the photospheric stage 
 should show the presence of CS lines of 
Ca\,II with P Cygni profile. We emphasize that the emission component 
is a signature 
for the confident distinguishing of CS lines from interstellar ones.

Another goal of the paper was the interpretation of CS lines detected 
in the spectrum of SN 1998S with a very dense wind. The modelling 
demonstrates that for the wind density $\omega=200$ and the black body 
spectrum of the supernova radiation the CS lines, especially Ca\,II, turn 
out to be too strong compared with observations. This controversy is 
resolved by assuming the existence of the ultraviolet excess with the 
relative flux fraction of about 6\%. We show that at the early stage 
$t<35$ d this ultraviolet excess can form owing to comptonization of 
the supernova radiation in the forward shock wave. At the later 
epoch the thermal radiation of the gas in the forward shock may 
contribute additionally, although this assumption requires a confirmation.

An observation of CS lines in SN IIP can be used to estimate the wind 
density. However, an example of SN~1998S shows that the equivalent 
width of the absorption depends non-monotonically on the wind density.
For $\omega<10$ we expect that equivalent width grows with the 
wind density, whereas in the region $\omega\sim10^2$ the equivalent width 
decreases with the wind density because of the ionization of metals in 
the wind by ultraviolet radiation produced by comptonization of optical 
photons on hot electrons of the forward shock. The use of the relation 
between the equivalent width of Ca\,II and $\omega$ for SN IIP 
is hampered by uncertainties 
related with the reduction factor of the ultraviolet radiation and 
with parameters of the turbulent velocity and the wind temperature.
By these reasons one hardly could measure the wind density with an accuracy 
better than factor of two. A wind clumpiness also affects the 
equivalent width. The effect of clumpiness is two-fold. First, the 
ionization decreases with the growing density. Therefore, for a given 
average column density the optical depth in clumpy case will be larger.
Second, for a given average column density of absorbing ions the 
equivalent width will be smaller, if the average number of clouds in the 
the line of sight is small, i.e., an order or less than unity. The 
expected modification of the line profile in this case is the decrease of 
the line depth because of incomplete covering of the photosphere by 
clouds. The effect of clumpiness will be especially apparent when 
profiles of the H and K lines of Ca\,II are compared. A similar relative 
intensities of these lines would evidence in favor of a saturation, while 
a shallow depth would indicate the clumpy structure of the wind with 
the average number of clouds in the line of sight of the order or less 
than unity.

We assumed that the wind is spherically-symmetric. In the case 
of asymmetric wind, e.g., equatorial wind, the emission component 
can become notably weaker than the absorption one, if the line of sight 
is close to the equatorial plane, or it can be stronger than the
absorption, if the line of sight is close to the polar axis. 
In the case of RSG strong deviations from spherical symmetry are 
unlikely, since SN IIP are single stars or components of wide binaries.
For example, Betelgeuse shows only weak deviations from spherical 
symmetry of its CS dusty envelope (Skinner et al. 1997) which indicates 
a quasi-spherical wind structure. Yet we should not rule out that 
in rare cases the SN IIP wind could be strongly asymmetric 
(SN 1987A is an example) because of close binary configuration. 
The line profile of Ca\,II 3934 \AA\ could be a valuable indicator of 
the asphericity of the wind outflow.

{}


\clearpage

\begin{figure}
\plotone{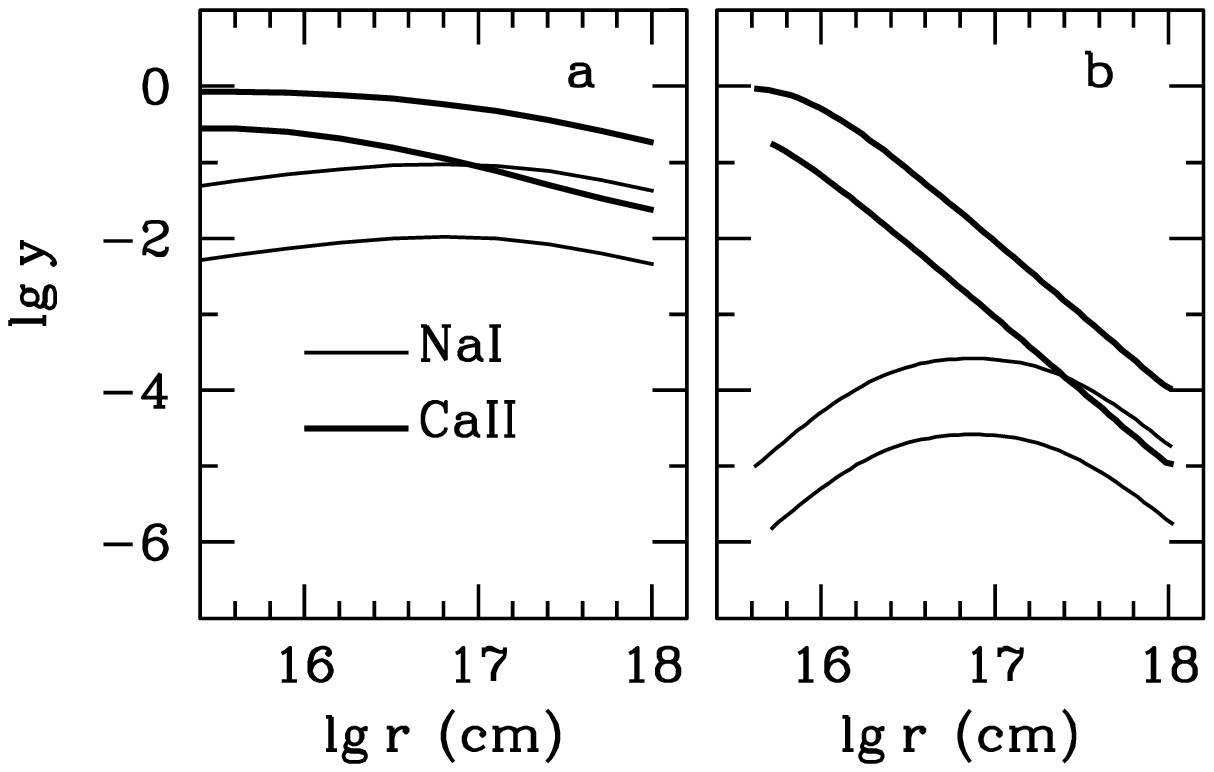}
\caption{
Fraction of ions Na\,I ({\em thin lines}) and Ca\,II ({\em thick lines})
in the wind before the supernova explosion (panel {\bf a}) and
50 days after (panel {\bf b}). Lower and upper line of each couple 
correspond to $\omega=1$ and $\omega=10$, respectively. 
}
 \end{figure}

\clearpage

\begin{figure}
\plotone{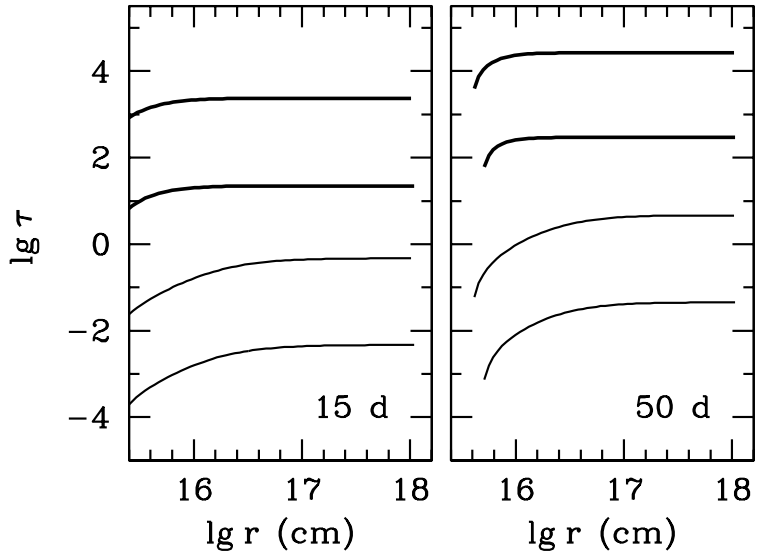}
\caption{
Optical depth of the wind in lines of Na\,I 5890 \AA\ ({\em thin lines}) 
and Ca\,II 3934 \AA\ ({\em thick lines}) integrated from the shock wave 
for two epochs. Lower and upper line of each couple 
correspond to $\omega=1$ and $\omega=10$, respectively. 
}
 \end{figure}

\clearpage

\begin{figure}
\plotone{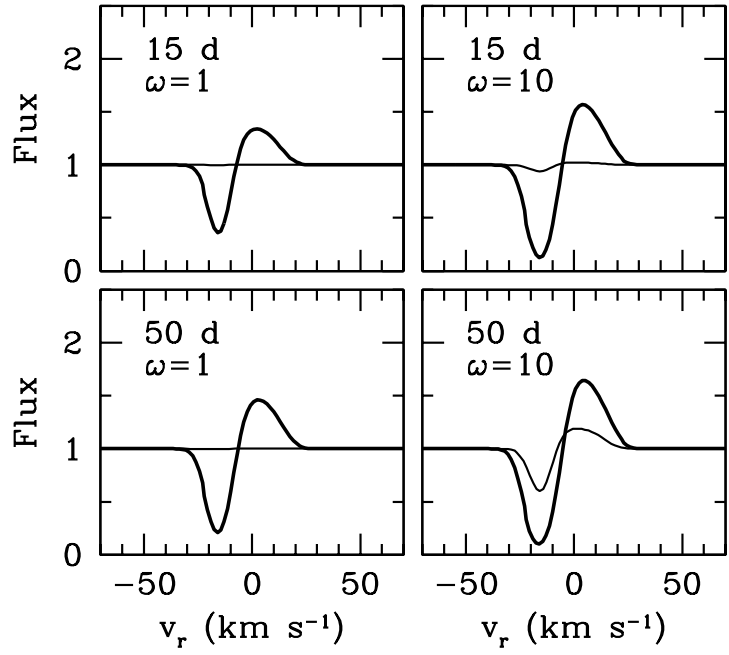}
\caption{
Model profiles of Na\,I 5890 \AA\ ({\em thin lines}) and 
Ca\,II 3934 \AA\ ({\em thick lines}) for two epochs and two 
values of the wind density.
}
 \end{figure}

\clearpage

\begin{figure}
\plotone{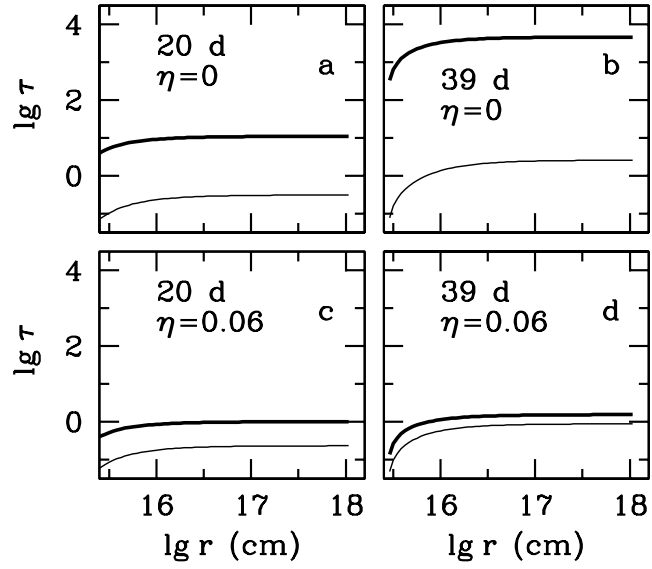}
\caption{
Optical depth of the wind in lines of Na\,I 5890 \AA\ ({\em thin lines})
and Ca\,II 3934 \AA\ ({\em thick lines}) in the model of the wind 
around SN 1998S for two epochs with ($\eta=0.06$) and without ($\eta=0$)
ultraviolet excess.
}
 \end{figure}

\clearpage

\begin{figure}
\plotone{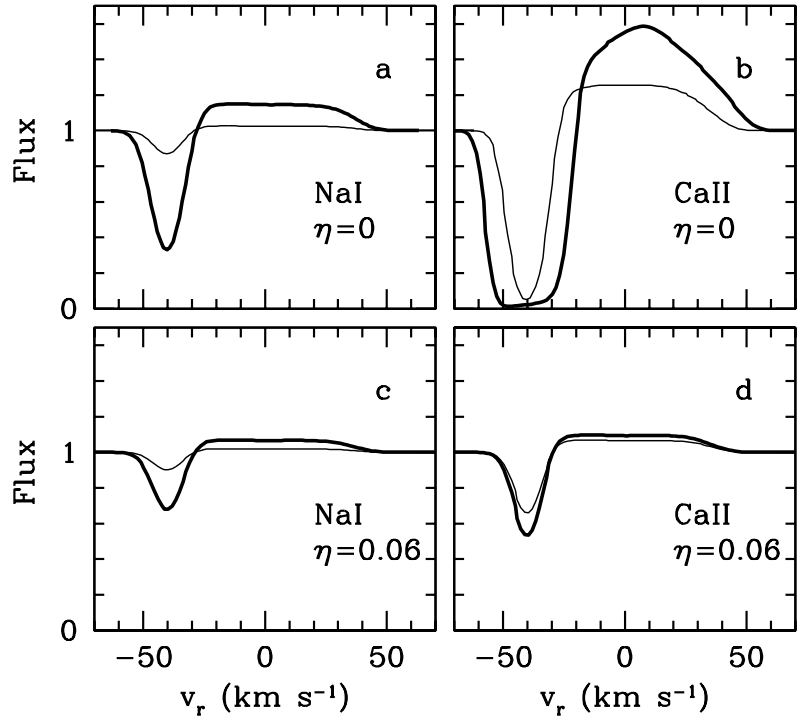}
\caption{
Model profiles of Na\,I 5890 \AA\ and Ca\,II 3934 \AA\ lines for two 
epochs without ($\eta=0$) and with ($\eta=0.06$) ultraviolet excess.
{\em Thin line} corresponds to the age of 20 d, while {\em thick line} 
corresponds to the age of 39 d.
}
 \end{figure}

\clearpage

\begin{figure}
\plotone{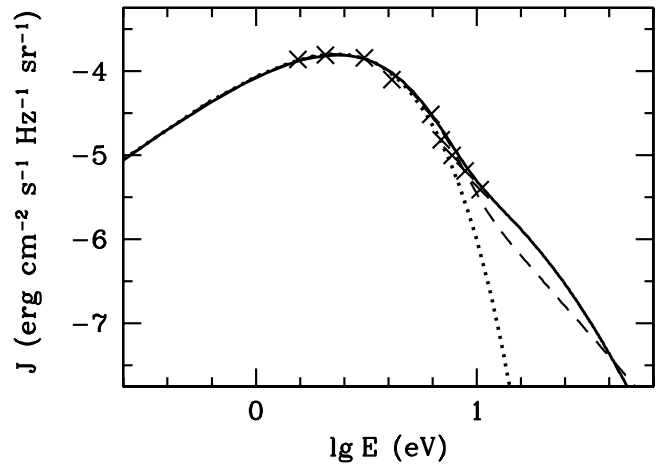}
\caption{
Comptonized spectrum ({\em solid line}) of black body radiation
({\em dotted line}) in comparison with the ultraviolet excess 
({\em dashed}) required to reproduce CS lines in SN 1998S. Crosses show 
the observed spectrum of SN 1998S on day 30 corrected for reddening.
}
 \end{figure}

\end{document}